\documentclass[conference]{IEEEtran}
\IEEEoverridecommandlockouts

\usepackage{cite}
\usepackage{amsmath,amssymb,amsfonts}
\usepackage{algorithmic}
\usepackage{graphicx}
\usepackage{textcomp}
\usepackage{xcolor}
\def\BibTeX{{\rm B\kern-.05em{\sc i\kern-.025em b}\kern-.08em
    T\kern-.1667em\lower.7ex\hbox{E}\kern-.125emX}}
\newcommand{\AuthorMark}[1]{\textsuperscript{#1}}

\usepackage{listings}
\usepackage{booktabs}
\usepackage{makecell}
\usepackage{multirow}
\usepackage{url}
\begin{document}

\title{QuCtrl-BELL: A Compiler-Driven Sub-Microsecond Feedback Control Stack for Scalable Trapped-Ion Quantum Experiments}

\author{\IEEEauthorblockN{Junpeng She\AuthorMark{1}, Ruoyu Yan\AuthorMark{1}, Zhizhen Qin\AuthorMark{3}, Zhanyu Li\AuthorMark{1}, \\ Zhongtao Shen\AuthorMark{3}, Zichao Zhou\AuthorMark{1,2}, Binxiang Qi\AuthorMark{1,*}, and Luming Duan\AuthorMark{1,2}}
    \IEEEauthorblockA{\AuthorMark{1}Center for Quantum Information, IIIS,
        Tsinghua University, Beijing 100084, China}
    \IEEEauthorblockA{\AuthorMark{2}Hefei National Laboratory, Hefei 230088, China}
    \IEEEauthorblockA{\AuthorMark{3}Department of Modern Physics,
        University of Science and Technology of China, Hefei 230026, China}
    \IEEEauthorblockA{\AuthorMark{*}Corresponding author: Binxiang Qi (qibinxiang@tsinghua.edu.cn)}
}

\maketitle
\begin{abstract}
    As trapped-ion quantum computing scales to larger qubit registers and more complex control protocols, classical control systems face a fundamental tradeoff: sub-microsecond board-level feedback requires tight hardware coupling, whereas maintainability and extensibility require clean, modular software abstractions. This paper presents QuCtrl-BELL (Bell), a compiler-driven software stack for trapped-ion quantum control. The design resolves this tradeoff by decoupling control flow---including loops, branches, and synchronization---from hardware state data. A Python-embedded domain-specific language (DSL) is lowered through a six-stage transpilation pipeline covering control flow graph (CFG) construction, static single-assignment (SSA) conversion, liveness analysis, and graph-coloring register allocation. The compiler generates deterministic distributed board-level programs and compact step-table data. A cross-board synchronization protocol supports feedback loops with latency below 700~ns without host intervention. Bell is deployed and evaluated on the QuCtrl-BELL platform (RISC-V + PXIe), demonstrating that a compiler-based infrastructure can provide programmability, deterministic timing, and modularity for scalable trapped-ion quantum control.
\end{abstract}
\begin{IEEEkeywords}
    trapped-ion quantum computing, quantum control software, domain-specific language, low-latency feedback, RISC-V
\end{IEEEkeywords}

\section{Introduction}

Trapped ions are among the most promising platforms for universal quantum computation, offering long coherence times, high-fidelity gates, and all-to-all qubit connectivity~\cite{Egan2021FaultTolerant,Niffenegger2020Integrated,Schaefer2018FastGates}. As experiments scale toward larger ion chains and more complex protocols, control demands grow correspondingly: signal channels expand from a handful to tens, timing logic evolves from linear sequences to nested loops and conditional branches, and real-time requirements tighten to sub-microsecond board-level feedback~\cite{Corcoles2020Challenges,Chu2024TITAN}. Recent trapped-ion experiments already span hundred-ion platforms and metropolitan ion-photon entanglement links~\cite{wu2023research,bigscale2025hamiltonian,Cui_Wang_Lai_Wang_Shi_Liu_Sun_Tian_Liang_Qi_et_al_2025}, further underscoring the importance of the electronic control stack for RF drive generation, detection gating, synchronization, and data acquisition.

Existing control software exhibits a persistent tradeoff. Systems implemented close to hardware registers can achieve low latency, but they are tightly coupled to a specific platform and difficult to reuse across experiments or backends. Systems that delegate timing logic to host-side scripts sacrifice determinism and cannot guarantee sub-microsecond latency. Modular laboratory-control suites such as Qudi~\cite{Binder2017Qudi} support instrument orchestration, but they do not define deterministic board-level timing semantics. A typical trapped-ion experiment involves DDS coherent output, TTL input/output, multi-board synchronization, parameter sweeps, and photon-detection-driven conditional branching---spanning both the slow host domain and the fast board-level execution domain.

Existing solutions address parts of this problem. ARTIQ/Sinara~\cite{Kasprowicz2020ARTIQ,Przywozki2023Sinara} offers mature real-time control within its hardware ecosystem, although cross-backend migration remains challenging~\cite{Erne2024ARTIQ}. QubiC~\cite{Xu2021QubiC,Xu2022QubiC,Xu2023QubiC2} and QICK~\cite{Ding2024QICK,DiGuglielmo2025QICK} demonstrate effective FPGA-based co-design for superconducting platforms. Pulselib~\cite{Alnas2025Pulselib} and TITAN~\cite{Chu2024TITAN} target hardware-agnostic pulse transpilation and distributed gate scheduling, respectively. More generally, compiler and interface abstractions such as realistic-hardware compiler design, OpenQASM, backend and pulse specifications, and high-level quantum DSLs have established the value of explicit intermediate representations and software/hardware interfaces~\cite{Chong2017Compiler,Cross2017OpenQASM,McKay2018Qiskit,Svore2018QSharp}, while other efforts focus on benchmarking, simulation acceleration, or network simulation rather than real-time hardware control~\cite{Chatterjee2025Benchmark,Li2024DDSim,DiAdamo2021QuNetSim}. Consequently, developing a unified path from experiment-level descriptions to distributed board-level execution that seamlessly integrates structured control flow with real-time feedback remains a critical frontier for advancing scalable quantum control.

This paper presents QuCtrl-BELL (Bell), a compiler-driven software stack for trapped-ion quantum control on the QuCtrl-BELL hardware platform. The central design principle is to compile \emph{control programs} (loops, branches, synchronization) separately from \emph{hardware state data} (DDS parameters, TTL masks). This separation yields a compact, reusable instruction stream while confining hardware-specific logic to well-defined code-generation and driver layers. The main contributions are as follows:
\begin{enumerate}
    \item A six-stage compilation pipeline that lowers a Python-embedded DSL through CFG construction, SSA conversion, liveness analysis, and graph-coloring register allocation into board-level programs and step-table data;
    \item A board-level real-time feedback architecture that realizes sub-microsecond feedback loops without host intervention;
    \item End-to-end deployment and evaluation on the QuCtrl-BELL trapped-ion control platform, validating compiler correctness and practical effectiveness.
\end{enumerate}

\section{System Design}

The design of Bell is derived from the control requirements of trapped-ion experiments and the strict execution constraints of distributed board-level hardware. This section is organized from physical constraints to software abstractions so that each compiler decision can be traced back to an experimental requirement. The discussion therefore proceeds from the experiment structure and hardware execution model to the architectural principles and programming abstractions imposed by them.

\subsection{Experimental Context and Hardware Constraints}

\begin{figure}[t]
    \centering
    \includegraphics[width=\columnwidth]{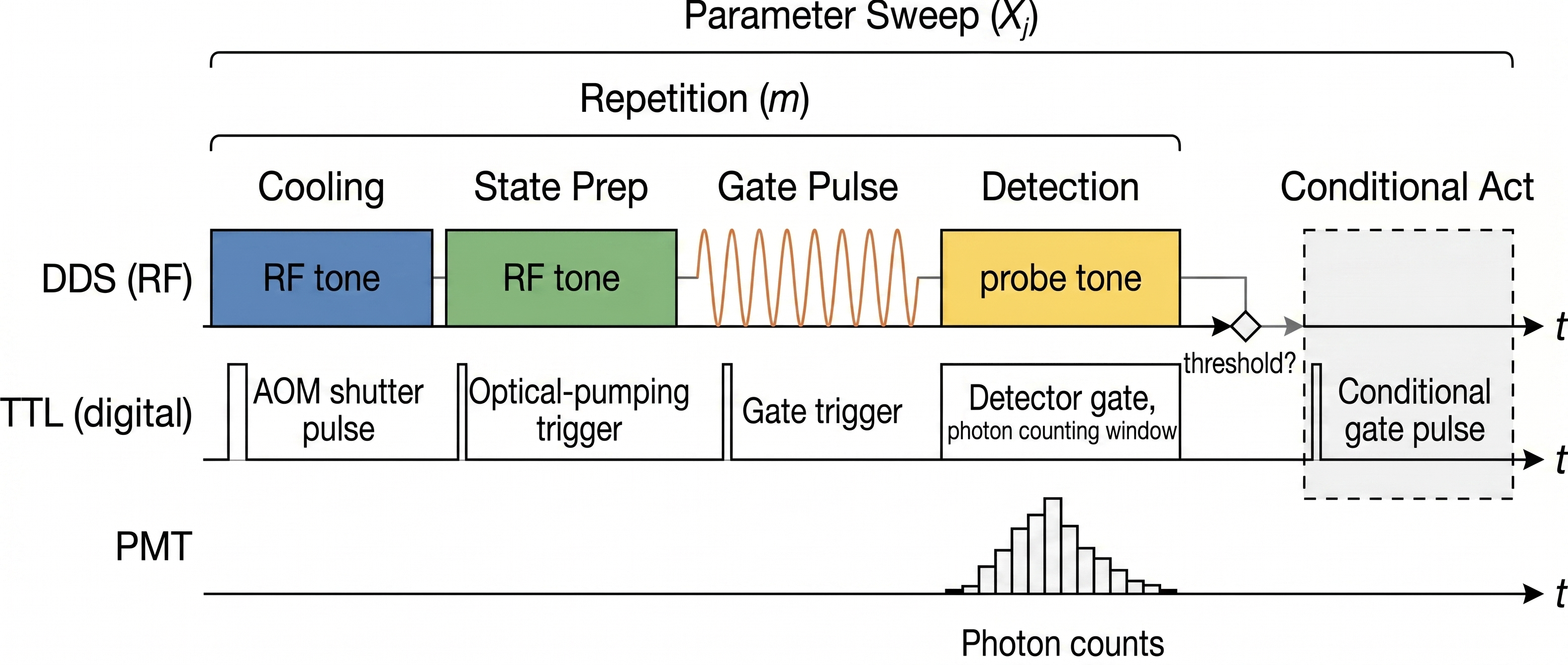}
    \caption{Canonical trapped-ion experiment sequence. Experiment comprises cooling, state preparation, a coherent gate pulse, and fluorescence detection. A threshold comparison on the photon count drives a conditional action at the board level. The shot repeats $m$ times; an outer loop sweeps parameter~$X_j$.}
    \label{fig:sequence}
\end{figure}

The control requirements addressed by Bell are most clearly exposed by the multi-phase sequence illustrated in Fig.~\ref{fig:sequence}. Each experimental shot always begins with Doppler \emph{cooling} to initialize the ion's motional state, followed by \emph{state preparation} (optical pumping) to project the internal state into a well-defined qubit basis. A coherent \emph{gate pulse}---driven by a DDS-generated RF tone whose frequency, amplitude, and phase are precisely programmed---then implements the desired quantum operation. During the subsequent \emph{detection} window, a TTL-gated photon-counting interval is opened on a PMT, producing a binary or accumulated count result. Based on this count, a \emph{conditional action} (e.g., a repump pulse) may then be applied in real time via board-level branching. The entire shot is repeated $m$ times for statistics, and an outer \emph{parameter sweep} loop varies a control parameter~$X_j$ (e.g., gate detuning or pulse duration) across the experiment.

From this sequence, four simultaneous requirements are imposed on the control stack: multi-channel synchronization across DDS and TTL boards, sub-microsecond feedback from detection to conditional branching, structured iteration with nested loops, and parametric variation across repeated shots.

\begin{figure}[t]
    \centering
    \includegraphics[width=1.0\columnwidth]{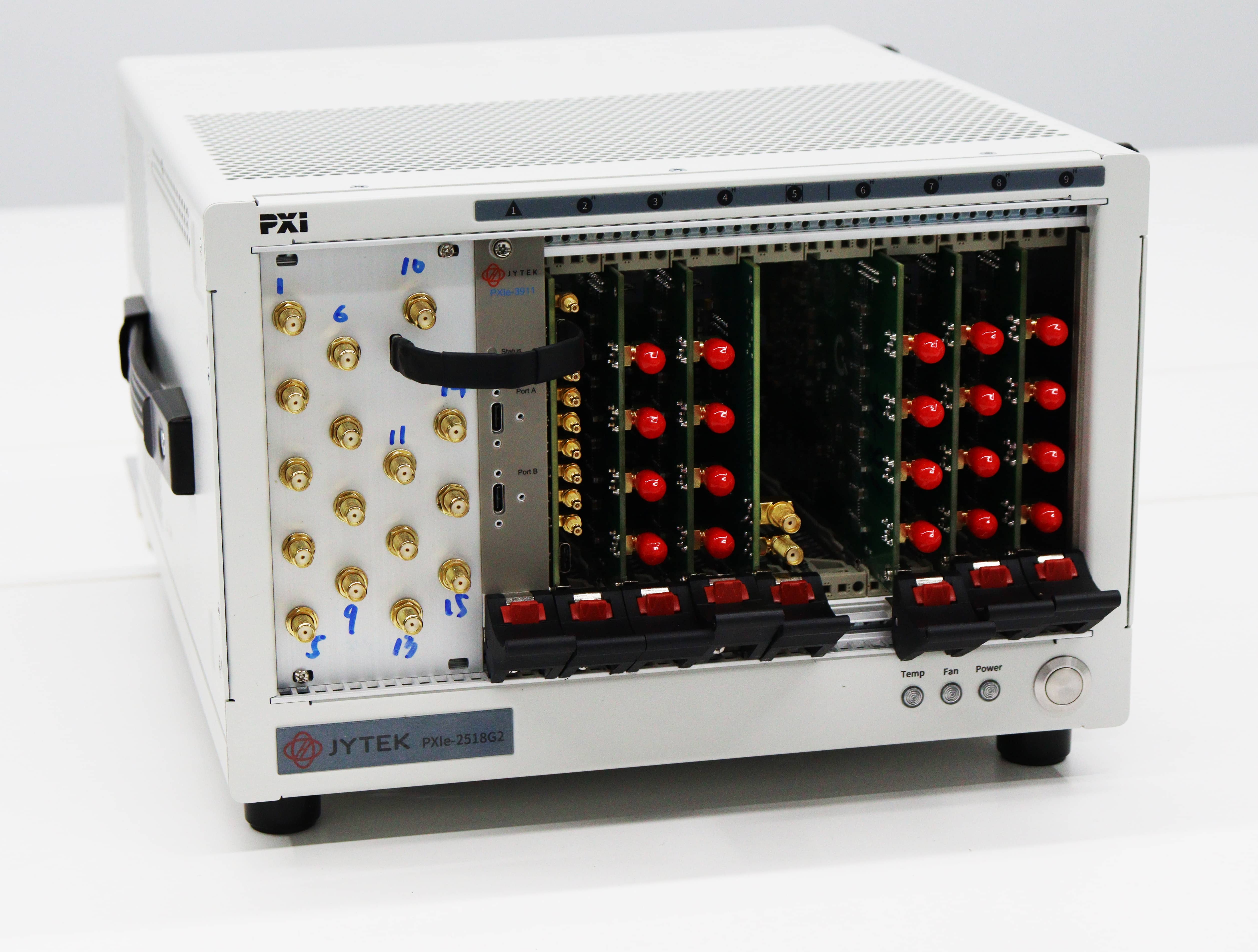}
    \caption{The QuCtrl-BELL hardware platform. All boards communicate via the PXIe backplane and the TCM star-bus broadcast network.}
    \label{fig:hardware}
\end{figure}

These requirements are mapped onto the QuCtrl-BELL hardware platform---Bell's primary deployment target---a distributed trapped-ion control system built on a 9-slot JYTEK PXIe chassis (Fig.~\ref{fig:hardware}). It is composed of DDS/AWG boards for coherent RF output, TTL/TDC boards for digital gating and photon counting, and a TCM (Trigger \& Clock Manager) for system-level clock distribution and star-topology broadcast. A single 9-slot chassis accommodates up to 24 DDS/AWG output channels and 32 TTL control channels, while its modular architecture ensures straightforward scalability to higher channel densities as the quantum system size increases. Deterministic synchronization and low-latency inter-device communication are recurring requirements in practical FPGA-based quantum electronics and distributed control fabrics~\cite{Stanco2022FPGA,Xu2025MultiFPGA}.

Each board hosts a local RISC-V processor onto which a \emph{step table}---an ordered list of hardware-state records---is preloaded; indices are fed into an execution queue at 4~ns clock resolution. Cross-board coordination is achieved through step-table alignment: step $i$ carries the same duration on every board, ensuring that all processors advance in lockstep. A fixed electronic delay below 100~ns has been measured for TTL readout and TCM broadcast. The remaining feedback latency arises from instruction-cycle overhead in branch execution on the local processors; its magnitude depends on the type and number of instructions preceding the branch. These constraints define the software boundary addressed by Bell: host participation cannot be tolerated on the critical feedback path, and compact control logic with predictable cross-board timing must be preserved.

\subsection{Core Design Principles}

In response to the hardware boundary conditions above, Bell is organized around three structural principles that reconcile high-level programmability with deterministic execution.

\textbf{P1 — Real-time-first synchronization semantics.} Board-level feedback (\texttt{read\_ttl}) and synchronization barriers are treated as first-class programming constructs rather than runtime-layer details. Their explicit representation at the language level ensures that timing boundaries are systematically reasoned about and that correct cross-board synchronization code is generated without hidden assumptions. Complementarily, \texttt{wait}/\texttt{resume} explicitly denotes the loss of real-time guarantees during host-side instrument coordination.

\textbf{P2 — Hardware decoupling through control/data separation.} Under this principle, compiled \emph{control programs} express loops, branches, synchronization points, and step-index management, whereas \emph{step tables} store the physical hardware parameters for each step (DDS frequency, amplitude, phase; TTL masks). Instructions reference step-table entries by index, and each entry is executed by the hardware at full clock resolution for its programmed duration. This separation keeps instruction streams compact regardless of iteration count, enables parameter sweeps by recompiling only the step table, and confines hardware specificity to the code-generation layers.

\textbf{P3 — Pipelined compilation with full observability.} Sequences are compiled through six ordered stages (detailed in Section~\ref{sec:pipeline}), each producing an inspectable intermediate artifact: node tree, CFG, SSA form, interference graph, register assignments, and final assembly. Observability is treated as a critical design requirement; transparent access to intermediate states enhances system robustness and streamlines debugging, thereby effectively mitigating the operational complexity of advanced trapped-ion experiments.

\subsection{Unified Software Stack and Programming Model}

\begin{figure}[t]
    \centering
    \includegraphics[width=0.8\columnwidth]{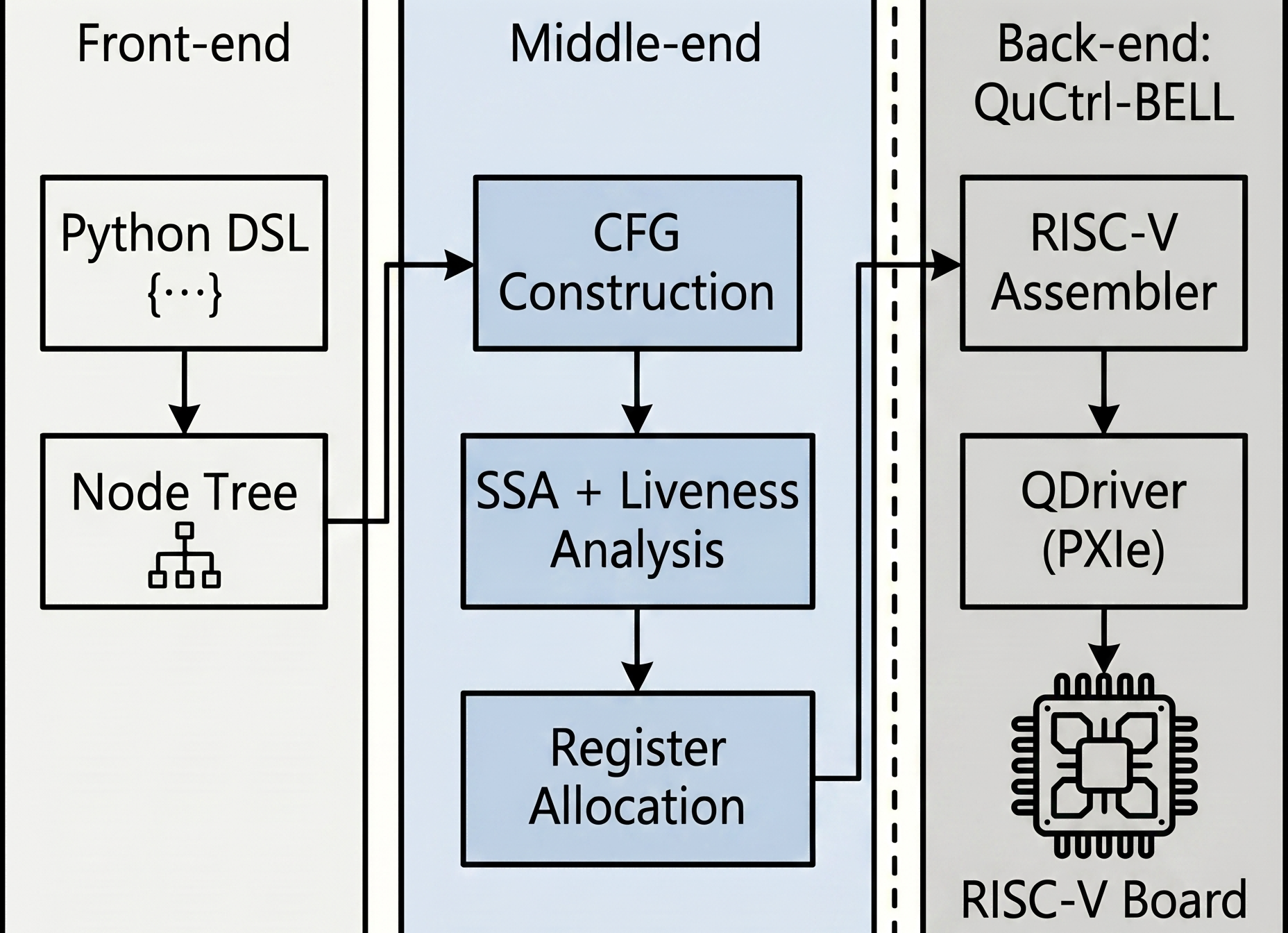}
    \caption{QuCtrl-BELL software architecture. The front-end (Python DSL and node tree) and middle-end (CFG, SSA, liveness, register allocation) are shared; the back-end below the dashed boundary provides hardware-targeted assembly generation and drivers.}
    \label{fig:architecture}
\end{figure}

On the basis of these principles, a unified three-tier compiler architecture is adopted (Fig.~\ref{fig:architecture}) so that the transition from high-level experiment description to hardware execution remains explicit.

The internal architecture divides cleanly into front-end, middle-end, and back-end. The \emph{front-end} captures user-written sequences as a typed node tree, preserving source-level structure. The \emph{middle-end} lowers this representation through CFG construction, SSA conversion, liveness analysis, and graph-coloring register allocation; these stages are hardware-independent and can therefore be shared across potential backends. Finally, the \emph{back-end} encodes the register-annotated representation into board-level programs and step-table data, targeting the QuCtrl-BELL platform via PCIe. By confining hardware-specific logic to this final tier, reuse is preserved above the backend boundary.

At the user level, this compiler stack is exposed through a Python-embedded domain-specific language (DSL) that provides three distinct abstraction layers without exposing board-level register operations:
\begin{itemize}
    \item \textbf{Channel level:} Device objects represent logical hardware channels and compose into named \emph{States} that capture multi-channel configurations.
    \item \textbf{Sequence level:} States are arranged temporally with structured control flow---including \texttt{loop}, \texttt{if\_}, \texttt{else\_}, and \texttt{while\_}---to build complete experiment programs.
    \item \textbf{Variable level:} Typed compile-time variables participate exclusively in board-side execution. Their values are never evaluated on the host, allowing threshold comparisons and loop counters to drive real-time feedback decisions directly within the compiled program.
\end{itemize}

A representative active-feedback fragment is shown in Listing~\ref{lst:feedback}. The sequence is expressed as a 20-iteration loop with a photon-count read and a conditional repump-cool path.

\begin{lstlisting}[caption={Active feedback in Bell's DSL. The entire feedback loop compiles to board-level instructions; no host involvement occurs during execution.},label={lst:feedback}]
with seq.loop(20):
    seq | Detect(1000)
    counts._ = seq.read_ttl(pmt)
    with seq.if_(counts._ < 5):
        seq | Repump(5000) | Cool(1000)
\end{lstlisting}

The \texttt{read\_ttl} primitive marks the only explicit inter-board synchronization boundary; the four-phase protocol discussed in Section~\ref{sec:feedback} is generated automatically by the compiler. Equivalent parameter sweeps can be expressed through a single \texttt{scan()} invocation without changing the sequence structure. Consequently, both sweep-oriented and feedback-oriented trapped-ion experiments can be expressed in Python while hardware register semantics remain abstracted from the user.


\section{Compilation Pipeline}\label{sec:pipeline}

\begin{figure}[t]
    \centering
    \includegraphics[width=0.95\columnwidth]{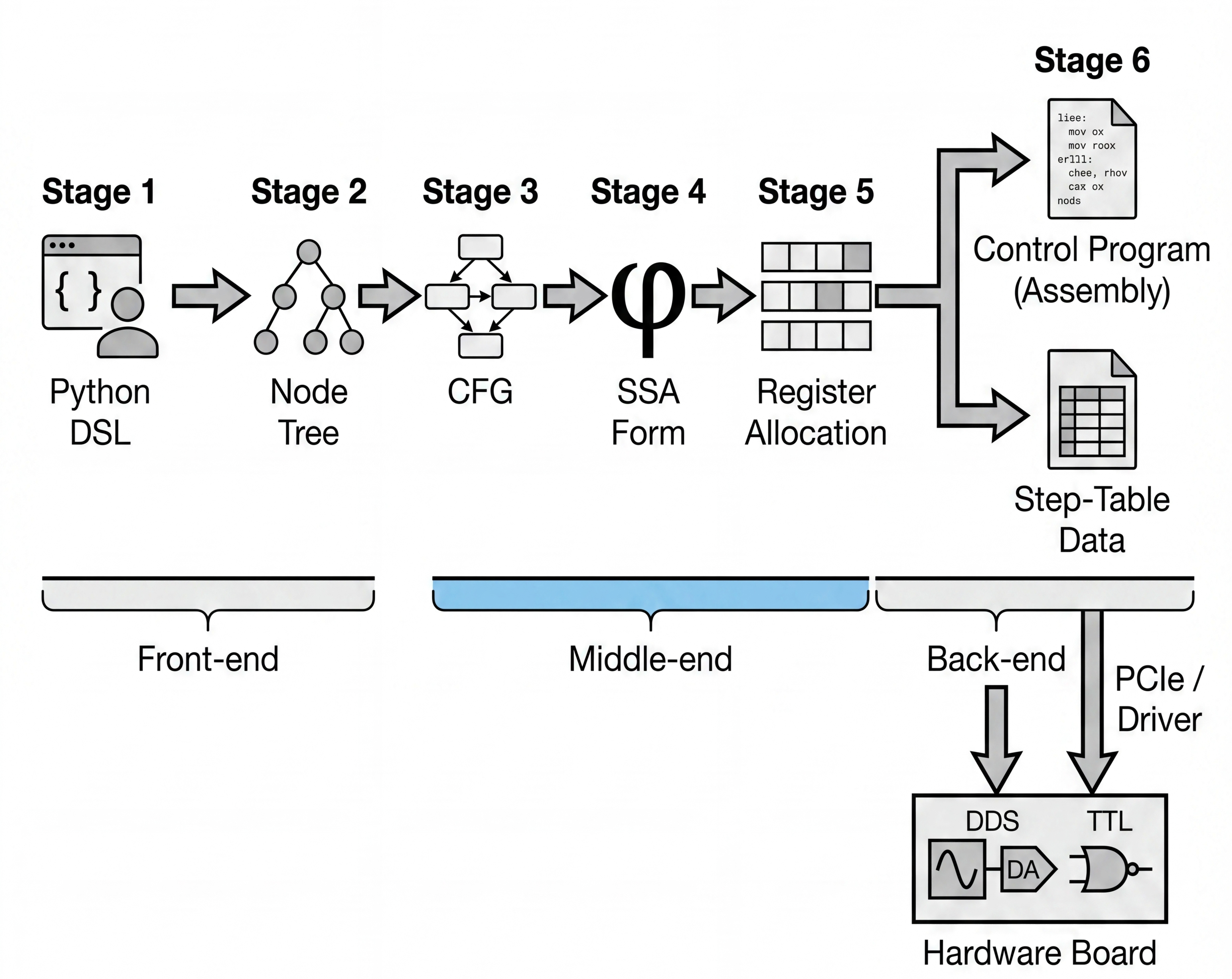}
    \caption{Bell's six-stage compilation pipeline. Stages 1--2 are the front-end, Stages 3--5 the middle-end, and Stage 6 emits the control program and step-table data sent to hardware via PCIe.}
    \label{fig:pipeline}
\end{figure}

To bridge high-level Python syntax and deterministic hardware execution, the compilation process is structured as a six-stage pipeline with explicit intermediate representations. User-written sequences are lowered from a typed node tree through CFG, SSA, liveness analysis, and graph-coloring register allocation into board-level symbolic assembly (Fig.~\ref{fig:pipeline}). In parallel, a dedicated data compiler generates per-board step-table files. Each stage emits an inspectable artifact, enabling debugging and validation to be performed at the abstraction level where an error is introduced. The remainder of this section is organized around the control/data decoupling strategy, the multi-stage lowering process, and the resulting real-time feedback implementation.

\subsection{Control/Data Decoupling Strategy}

In accordance with P2, two complementary artifacts are generated by the compiler. The \emph{control program} contains loop, branch, synchronization, and step-index management instructions. The \emph{step table} stores the per-step physical hardware state: DDS channel frequencies, amplitudes, and phases; TTL output and input masks; and step durations. Instructions reference step-table entries by index, and each entry is executed by the hardware at full 4~ns clock resolution.

Three system-level benefits are obtained from this separation. First, \emph{compactness}: a 50-iteration loop over two hardware states generates two step-table entries and a small number of loop instructions, rather than 100 inlined configurations. Second, \emph{parametric reuse}: parameter sweeps recompile only the step table while the control program remains unchanged, directly reducing per-point compilation overhead. Third, \emph{modularity}: the control-program format and step-table schema remain structurally independent of the code-generation layer, so hardware-specific logic is confined to a well-defined substitution boundary and future backend adaptation is simplified.

\subsection{Multi-Stage Lowering Process}

The lowering flow is organized across front-end, middle-end, and back-end transformations, and each transition is mediated by an explicit intermediate representation (IR). The node tree preserves typed DSL intent, the CFG exposes control transfer and synchronization structure, SSA and liveness information expose data dependencies, and the register-annotated CFG provides the machine-oriented view from which board programs are emitted.

The \emph{front-end} (Stages 1--2) records each DSL call as a typed node carrying source-location metadata, producing an ordered node tree. This representation preserves user intent while remaining independent of any specific hardware instruction set. The \emph{middle-end} (Stages 3--5) first lowers this tree into a CFG of basic blocks: loops become back edges, conditionals introduce branch/merge block pairs, and explicit transfer constructs add direct edges. Because all execution paths are made explicit in this IR, cross-branch variable analysis, synchronization-boundary determination, and dead-code detection become possible.

The CFG is then converted to SSA form via the dominance-frontier algorithm, with $\phi$-functions inserted at merge points to reconcile variable versions from different predecessors. Liveness analysis is subsequently reduced to a standard backward dataflow computation, producing per-variable live ranges and an interference graph. A graph-coloring register allocator finally maps program variables to the board processor's constrained register set; when register pressure exceeds capacity, spill code is inserted automatically, with loop-carried variables prioritized to minimize overhead on hot paths.

The \emph{back-end} (Stage 6) encodes the register-annotated CFG into board-loadable machine words. Custom coordination instructions are used for step-index management and inter-board data reception, while standard RISC-V instructions are used for arithmetic and control flow. All instruction encodings are validated by the unit test suite before deployment.

\subsection{Real-Time Feedback Implementation}\label{sec:feedback}

\begin{figure}[t]
    \centering
    \includegraphics[width=0.8\columnwidth]{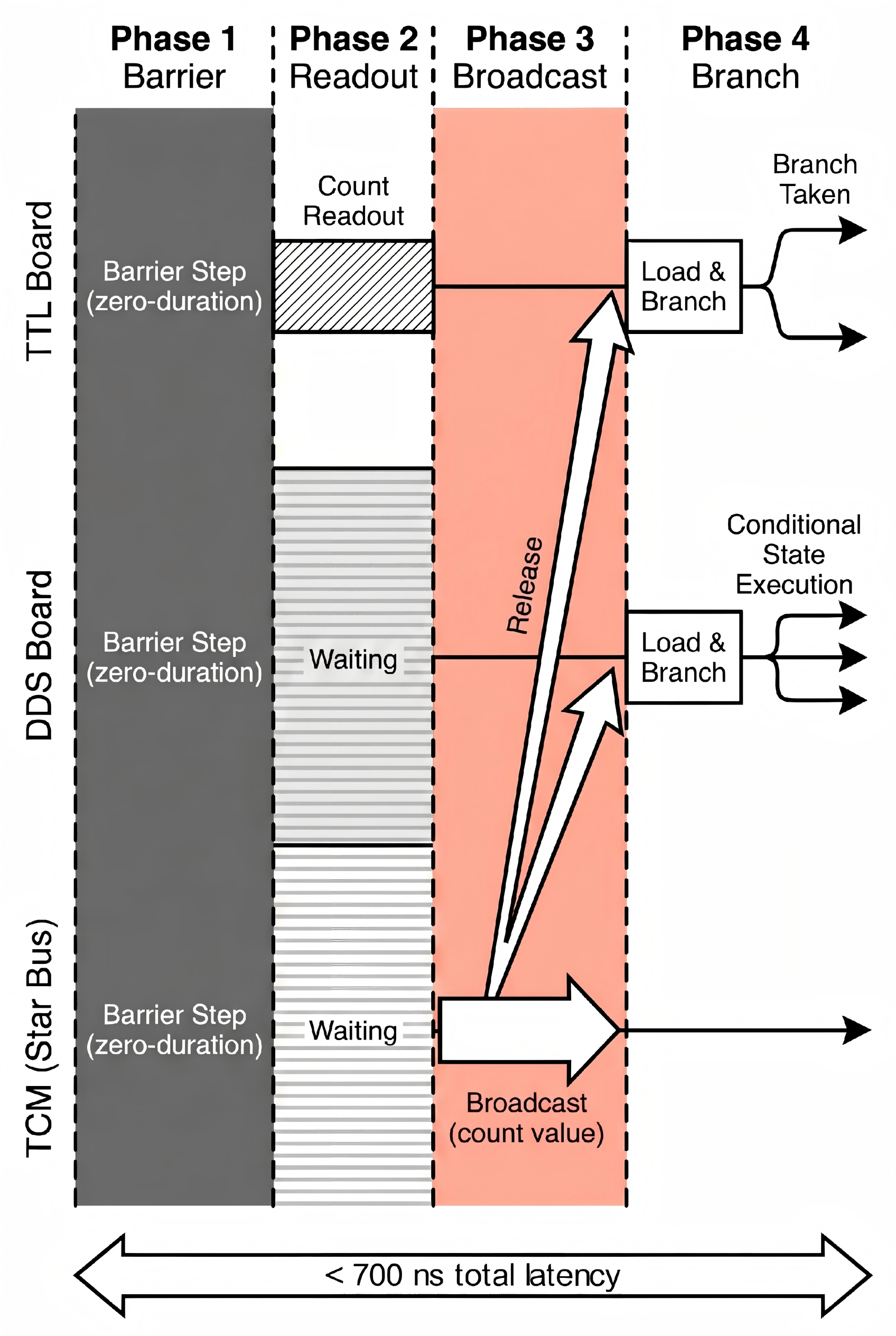}
    \caption{Cross-board synchronization for \texttt{read\_ttl}: (1) barrier, (2) TTL photon-count readout, (3) TCM broadcast release, and (4) branch execution on each board. Total latency is below 700~ns without host round-trips.}
    \label{fig:sync}
\end{figure}

Among the generated constructs, the \texttt{read\_ttl} operation is the most timing-critical because a photon count acquired on the TTL board must be made visible to all participating boards before a branch can be resolved. The operation is implemented through the four-phase protocol illustrated in Fig.~\ref{fig:sync}: (1)~a \emph{barrier step} halts all board processors at the synchronization boundary; (2)~the TTL board reads the photon count into a local register; (3)~the TTL board broadcasts the count value via the TCM star bus, atomically releasing all waiting boards; (4)~each board loads the broadcast value and executes its compiled conditional branch. Each phase is fixed at compile time, and no host interaction is introduced on the critical timing path. The measured end-to-end latency remains below 700~ns.

\section{Evaluation and Experimental Results}

The design claims of Bell are evaluated from three complementary perspectives: positioning against representative quantum-control frameworks, host-side compilation throughput with board-level logic verification, and memory efficiency obtained from control/data separation.

\subsection{Benchmarking against Existing Frameworks}

Bell is positioned against representative quantum control systems in Table~\ref{tab:comparison}. To the best of our knowledge, it is the only system that combines a full compiler pipeline (CFG, SSA, register allocation), explicit control/data separation, and sub-microsecond board-level feedback. Among feedback-capable systems, ARTIQ is optimized for the Sinara ecosystem, QubiC emphasizes parametric FPGA feedback, and QICK provides branching through tProc, with broader control-flow support still under development. Other systems prioritize different goals, including subroutine reuse~\cite{Dalvi2024Subroutine}, hardware-agnostic pulse scheduling~\cite{Alnas2025Pulselib}, and distributed zone routing~\cite{Chu2024TITAN}, rather than closed-loop board-level feedback.

\begin{table}[t]
    \caption{Comparison with Representative Quantum Control Systems}
    \label{tab:comparison}
    \centering
    \resizebox{\columnwidth}{!}{%
        \begin{tabular}{@{}lcccc@{}}
            \toprule
            \textbf{System}                                      &
            \makecell[c]{\textbf{Prog.}                                                                                                                      \\\textbf{Model}} &
            \makecell[c]{\textbf{Compiler}                                                                                                                   \\\textbf{Pipeline}} &
            \makecell[c]{\textbf{Ctrl/Data}                                                                                                                  \\\textbf{Sep.}} &
            \makecell[c]{\textbf{Board-Level}                                                                                                                \\\textbf{Feedback}} \\
            \midrule
            ARTIQ \cite{Kasprowicz2020ARTIQ,Przywozki2023Sinara} & Py subset       & AST$\to$LLVM        & No           & $\sim$3\,$\mu$s                    \\
            QubiC \cite{Xu2021QubiC,Xu2023QubiC2}                & Py API          & Parametric          & Partial      & Yes                                \\
            QICK \cite{Ding2024QICK,DiGuglielmo2025QICK}         & Py API          & Macro-ASM           & No           & Limited                            \\
            Dalvi et al.\ \cite{Dalvi2024Subroutine}             & Gate sub.       & Subroutine          & Partial      & No                                 \\
            pulselib \cite{Alnas2025Pulselib}                    & Pulse graph     & Graph tx.           & ---          & No                                 \\
            TITAN \cite{Chu2024TITAN}                            & Gate IR         & Scheduling          & ---          & No                                 \\
            \midrule
            \textbf{QuCtrl-BELL (Ours)}                          & \textbf{Py DSL} & \textbf{CFG/SSA/RA} & \textbf{Yes} & $\boldsymbol{<}$\,\textbf{700\,ns} \\
            \bottomrule
        \end{tabular}%
    }
\end{table}


\subsection{Pipeline Throughput and Logic Verification}

Compilation throughput and execution determinism are characterized jointly because both properties are required for practical deployment. Six representative programs of increasing complexity are compiled to generate the metrics summarized in Table~\ref{tab:compilation}. All programs compile entirely on the host in under 4~ms wall-clock time, with each pipeline stage (CFG construction, SSA conversion, liveness analysis, register allocation) completing in the single-digit millisecond range---negligible relative to the multi-second shot durations of typical trapped-ion experiments.

Several structural observations follow from Table~\ref{tab:compilation}. First, assembly instruction counts remain in the tens even for programs with nested feedback loops, which is consistent with P2 because per-step hardware parameters are excluded from the instruction stream. Second, step-table size scales with the number of \emph{distinct} hardware configurations rather than with iteration count, anticipating the memory analysis in Section~\ref{sec:compactness}. Third, register allocation completes without spills for all tested programs; spill code is inserted automatically only when simultaneously live variables exceed the register file capacity.

\begin{figure}[t]
    \centering
    \includegraphics[width=\columnwidth]{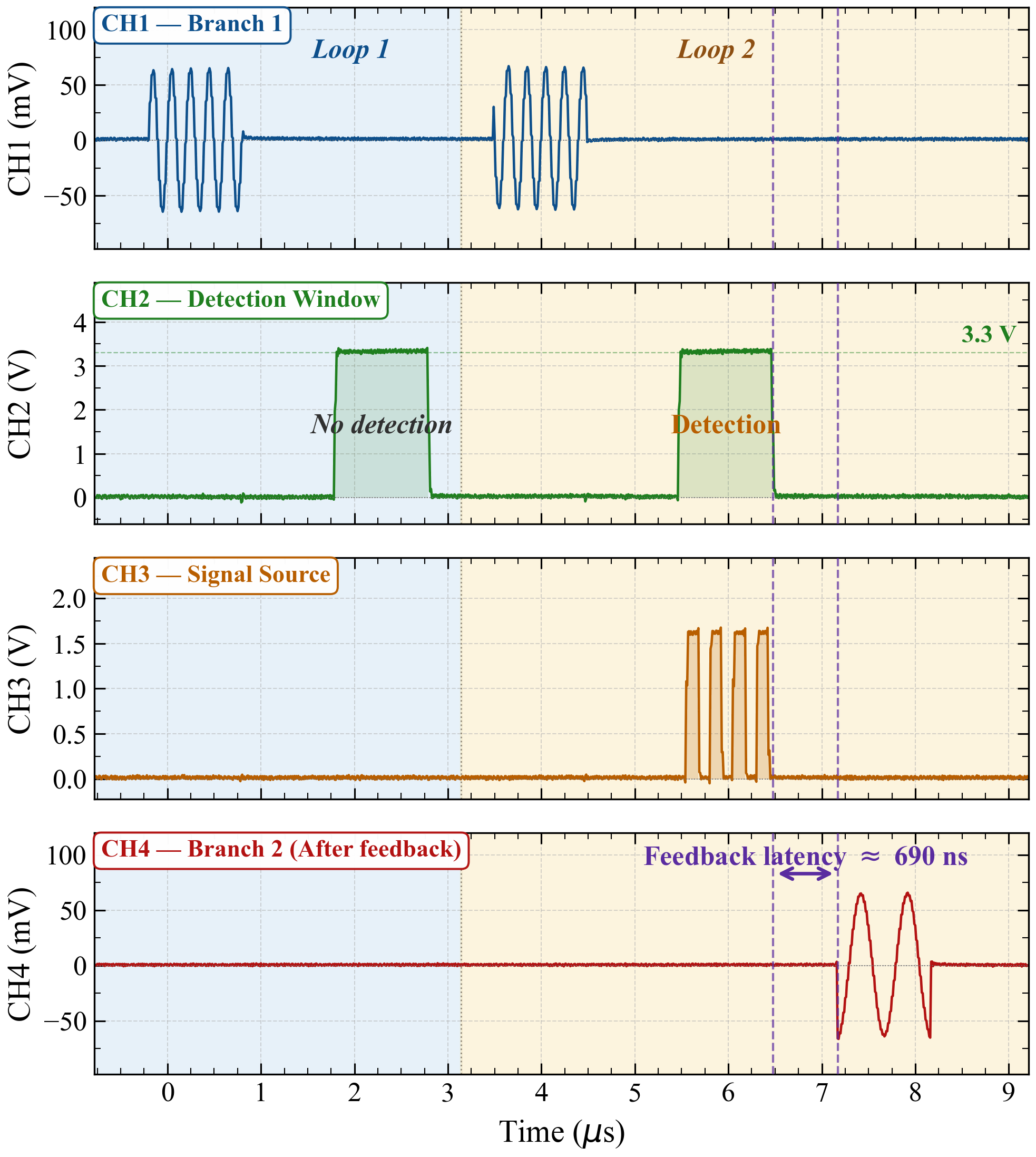}
    \caption{Oscilloscope trace of one board-level feedback cycle. \textbf{CH1}: repeated DDS test-gate output. \textbf{CH2}: TTL detection window, with the second pulse indicating an active event. \textbf{CH3}: photon-counter readback. \textbf{CH4}: DDS output triggered by the feedback decision. The marked delay from the CH2 falling edge to the CH4 rising edge is 690~ns, confirming sub-700~ns response without host intervention.}
    \label{fig:latency_scope}
\end{figure}

Logic verification is supported at both compiler and hardware levels. The generated control flow remains inspectable through the CFG, SSA, and register-allocation artifacts discussed in Section~\ref{sec:pipeline}, while runtime determinism is confirmed directly on the hardware platform. The \texttt{read\_ttl} feedback operation introduces approximately 700~ns of deterministic board-level synchronization overhead per invocation; for feedback-free sequences, execution overhead is negligible because board processors execute asynchronously ahead of the hardware timeline~\cite{Corcoles2020Challenges}. Fig.~\ref{fig:latency_scope} shows an oscilloscope capture of a complete feedback loop on the QuCtrl-BELL platform: the detection window (CH2) closes, and the conditional DDS output (CH4) rises 690~ns later, confirming the sub-700~ns target without host intervention on the critical path.

\begin{table}[t]
    \caption{Compilation Characteristics of Representative Programs}
    \label{tab:compilation}
    \centering
    \resizebox{\columnwidth}{!}{%
        \begin{tabular}{lcccc}
            \toprule
            \textbf{Program}     & \textbf{CFG BBs} & \textbf{SSA Vars} & \textbf{Asm Instrs} & \textbf{ST Entries} \\
            \midrule
            Simple pulse         & 6                & 0                 & 16                  & 4                   \\
            Variable readout     & 8                & 4                 & 25                  & 4                   \\
            Active feedback      & 10               & 4                 & 30                  & 6                   \\
            Nested loop          & 12               & 5                 & 34                  & 4                   \\
            Multi-var feedback   & 11               & 11                & 42                  & 7                   \\
            While-loop threshold & 7                & 4                 & 22                  & 4                   \\
            \bottomrule
        \end{tabular}%
    }
    \vspace{-6pt}
\end{table}

\subsection{Memory Efficiency via Structural Compactness}\label{sec:compactness}

The memory consequence of P2 is quantified by comparing Bell's step-table size against the hardware-configuration records that would be required by a naive inlining strategy. In an inlined execution model, each loop iteration over $k$ distinct hardware states produces $k$ configuration records transferred to the board, totaling $n \times k$ records over $n$ iterations. Under Bell's representation, only $k$ step-table entries are retained regardless of $n$; the loop itself is represented once in the control program, and the step table is indexed at runtime without redundant duplication.

Table~\ref{tab:compactness} quantifies this for the six benchmark programs, using per-program iteration counts representative of typical trapped-ion experiment loops.

\begin{table}[t]
    \caption{Step-Table Compactness: Bell vs.\ Naive Inlining}
    \label{tab:compactness}
    \centering
    \resizebox{\columnwidth}{!}{%
        \begin{tabular}{lcccc}
            \toprule
            \textbf{Program}     & \textbf{Iterations} & \textbf{ST Entries} & \textbf{Naive Configs} & \textbf{Reduction} \\
            \midrule
            Simple pulse         & 100                 & 4                   & 400                    & 100$\times$        \\
            Variable readout     & 50                  & 4                   & 200                    & 50$\times$         \\
            Active feedback      & 20                  & 6                   & 120                    & 20$\times$         \\
            Nested loop          & 1000                & 4                   & 4000                   & 1000$\times$       \\
            Multi-var feedback   & 100                 & 7                   & 700                    & 100$\times$        \\
            While-loop threshold & ---                 & 4                   & ---                    & ---                \\
            \bottomrule
        \end{tabular}%
    }
    \vspace{-6pt}
\end{table}

For active-feedback and loop-sweep programs---the dominant workloads in trapped-ion experiments---the reduction in step-table size directly lowers board memory pressure and upload bandwidth in proportion to total iteration count. For parameter sweeps, only the step table is recompiled per scan point while the control program is held constant, eliminating per-point instruction-stream regeneration and making sweep overhead largely independent of program structural complexity. This behavior provides a direct quantitative validation of P2.

\section{Conclusion and Outlook}

This work demonstrates that compiler techniques can reconcile high-level programmability with hardware determinism in trapped-ion quantum control. By separating control programs from hardware state data, Bell supports expressive experiment-level programming through a Python DSL while preserving deterministic board-level execution with sub-microsecond feedback latency.

Several extensions remain. First, each \texttt{read\_ttl} event introduces roughly 700~ns of synchronization overhead. In rapid-feedback scenarios such as multi-round quantum error correction, this overhead may become a throughput bottleneck. Hardware-assisted pipelining of broadcast events is therefore a promising direction for amortizing this cost. Second, camera-based (IMG) detection feedback, although already supported at the hardware level, has not been integrated into the compiler; extending the scalar \texttt{read\_ttl} abstraction to image-derived multi-qubit readout remains a near-term objective. Third, Bell's modular architecture---with hardware specificity confined to the assembler, step-table encoder, and driver---is intended to support multi-backend portability; adaptation to RFSoC-based platforms would provide a direct empirical test of this design boundary.

Deployed and validated on the QuCtrl-BELL platform, Bell provides a modular, programmable, and low-latency infrastructure for next-generation large-scale trapped-ion quantum control systems.

\section*{Acknowledgment}

This work was supported by Quantum Science and Technology-National Science and Technology Major Project (2021ZD0301601). The authors used GitHub Copilot for language editing and grammar enhancement throughout the article, and Nano-Banana for refining selected figures excluding photographs of physical devices or experimental setups. All content was reviewed and edited by the authors, who take full responsibility for the final work.

\bibliographystyle{IEEEtran}
\bibliography{IEEEabrv,refs}

\begin{thebibliography}{10}
\providecommand{\url}[1]{#1}
\csname url@samestyle\endcsname
\providecommand{\newblock}{\relax}
\providecommand{\bibinfo}[2]{#2}
\providecommand{\BIBentrySTDinterwordspacing}{\spaceskip=0pt\relax}
\providecommand{\BIBentryALTinterwordstretchfactor}{4}
\providecommand{\BIBentryALTinterwordspacing}{\spaceskip=\fontdimen2\font plus
\BIBentryALTinterwordstretchfactor\fontdimen3\font minus \fontdimen4\font\relax}
\providecommand{\BIBforeignlanguage}[2]{{%
\expandafter\ifx\csname l@#1\endcsname\relax
\typeout{** WARNING: IEEEtran.bst: No hyphenation pattern has been}%
\typeout{** loaded for the language `#1'. Using the pattern for}%
\typeout{** the default language instead.}%
\else
\language=\csname l@#1\endcsname
\fi
#2}}
\providecommand{\BIBdecl}{\relax}
\BIBdecl

\bibitem{Egan2021FaultTolerant}
L.~Egan \emph{et~al.}, ``Fault-tolerant control of an error-corrected qubit,'' \emph{Nature}, vol. 598, no. 7880, pp. 281--286, Oct. 2021.

\bibitem{Niffenegger2020Integrated}
R.~J. Niffenegger \emph{et~al.}, ``Integrated multi-wavelength control of an ion qubit,'' \emph{Nature}, vol. 586, no. 7830, pp. 538--542, Oct. 2020.

\bibitem{Schaefer2018FastGates}
V.~M. Sch{\"a}fer, C.~J. Ballance, K.~Thirumalai \emph{et~al.}, ``Fast quantum logic gates with trapped-ion qubits,'' \emph{Nature}, vol. 555, no. 7694, pp. 75--78, Mar. 2018.

\bibitem{Corcoles2020Challenges}
A.~D. C{\'o}rcoles \emph{et~al.}, ``Challenges and opportunities of near-term quantum computing systems,'' \emph{Proceedings of the IEEE}, vol. 108, no.~8, pp. 1338--1352, Aug. 2020.

\bibitem{Chu2024TITAN}
C.~Chu, Z.~Fu, Y.~Xu \emph{et~al.}, ``{TITAN}: A fast and distributed large-scale trapped-ion {NISQ} computer,'' in \emph{Proc. 61st ACM/IEEE Design Autom. Conf. (DAC)}, Nov. 2024, pp. 1--6.

\bibitem{wu2023research}
Y.-K. Wu and L.-M. Duan, ``Research progress of ion trap quantum computing,'' \emph{Acta Physica Sinica}, vol.~72, no.~23, 2023.

\bibitem{bigscale2025hamiltonian}
S.-A. Guo, Y.-K. Wu, J.~Ye, L.~Zhang, Y.~Wang, W.-Q. Lian, R.~Yao, Y.-L. Xu, C.~Zhang, Y.-Z. Xu, B.-X. Qi, P.-Y. Hou, L.~He, Z.-C. Zhou, and L.-M. Duan, ``Hamiltonian learning for 300 trapped-ion qubits with long-range couplings,'' \emph{Science Advances}, vol.~11, no.~5, p. eadt4713, 2025.

\bibitem{Cui_Wang_Lai_Wang_Shi_Liu_Sun_Tian_Liang_Qi_et_al_2025}
Z.-B. Cui, Z.-Q. Wang, P.-C. Lai, Y.~Wang, J.-X. Shi, P.-Y. Liu, Y.-D. Sun, Z.-C. Tian, Y.-B. Liang, B.-X. Qi, Y.-Y. Huang, Z.-C. Zhou, Y.-K. Wu, Y.~Xu, L.-M. Duan, and Y.-F. Pu, ``Metropolitan-scale ion-photon entanglement via a quantum network node with hybrid multiplexing enhancements,'' \emph{Nature Communications}, vol.~17, no.~1, p. 697, Dec. 2025.

\bibitem{Binder2017Qudi}
J.~M. Binder, A.~Stark, N.~Tomek, and J.~Scheuer, ``{Qudi}: A modular python suite for experiment control and data processing,'' \emph{SoftwareX}, vol.~6, pp. 85--90, 2017.

\bibitem{Kasprowicz2020ARTIQ}
G.~Kasprowicz \emph{et~al.}, ``{ARTIQ} and {Sinara}: Open software and hardware stacks for quantum physics,'' in \emph{OSA Quantum 2.0 Conf.}, 2020, p. QTu8B.14.

\bibitem{Przywozki2023Sinara}
T.~Przyw{\'o}zki \emph{et~al.}, ``{Sinara} and {ARTIQ}: Open-source ion-trapping control system,'' in \emph{Proc. IEEE Int. Conf. Quantum Comput. Eng. (QCE)}, Sep. 2023, pp. 294--295.

\bibitem{Erne2024ARTIQ}
J.~Erne, ``Development of {ARTIQ}-based control software for quantum gas experiments,'' Master's thesis, ETH Zurich, Zurich, Switzerland, 2024.

\bibitem{Xu2021QubiC}
Y.~Xu \emph{et~al.}, ``{QubiC}: An open-source {FPGA}-based control and measurement system for superconducting quantum information processors,'' \emph{IEEE Transactions on Quantum Engineering}, vol.~2, pp. 1--11, 2021.

\bibitem{Xu2022QubiC}
Y.~Xu, G.~Huang, J.~Balewski \emph{et~al.}, ``Automatic qubit characterization and gate optimization with {QubiC},'' \emph{ACM Transactions on Quantum Computing}, vol.~4, pp. 3:1--3:12, 2022.

\bibitem{Xu2023QubiC2}
Y.~Xu, G.~Huang, N.~Fruitwala \emph{et~al.}, ``{QubiC} 2.0: An extensible open-source qubit control system capable of mid-circuit measurement and feed-forward,'' arXiv:2309.10333, 2023.

\bibitem{Ding2024QICK}
C.~Ding \emph{et~al.}, ``Experimental advances with the {QICK} ({Quantum Instrumentation Control Kit}) for superconducting quantum hardware,'' \emph{Physical Review Research}, vol.~6, no.~1, p. 013305, Mar. 2024.

\bibitem{DiGuglielmo2025QICK}
G.~{Di Guglielmo} \emph{et~al.}, ``End-to-end workflow for machine-learning-based qubit readout with {QICK} and hls4ml,'' \emph{IEEE Transactions on Quantum Engineering}, vol.~6, pp. 1--10, 2025.

\bibitem{Alnas2025Pulselib}
J.~Alnas, A.~S. Dalvi, and K.~R. Brown, ``Towards a pulse-level intermediate representation for diverse quantum control systems,'' in \emph{Proc. IEEE Int. Conf. Quantum Comput. Eng. (QCE)}, Aug. 2025, pp. 448--458.

\bibitem{Chong2017Compiler}
F.~T. Chong, D.~Franklin, and M.~Martonosi, ``Programming languages and compiler design for realistic quantum hardware,'' \emph{Nature}, vol. 549, pp. 180--187, 2017.

\bibitem{Cross2017OpenQASM}
A.~W. Cross, L.~S. Bishop, J.~A. Smolin, and J.~M. Gambetta, ``{OpenQASM}: Open quantum assembly language,'' arXiv:1707.03429, 2017.

\bibitem{McKay2018Qiskit}
D.~C. McKay, T.~Alexander, L.~Bello \emph{et~al.}, ``{Qiskit} backend specifications for {OpenQASM} and {OpenPulse} experiments,'' arXiv:1809.03452, 2018.

\bibitem{Svore2018QSharp}
K.~Svore, A.~Geller, M.~Troyer \emph{et~al.}, ``{Q\#}: Enabling scalable quantum computing and development with a high-level {DSL},'' in \emph{Proc. Real World Domain Specific Languages Workshop (RWDSL)}, 2018, pp. 1--10.

\bibitem{Chatterjee2025Benchmark}
A.~Chatterjee, S.~Rappaport, A.~Giri \emph{et~al.}, ``A comprehensive cross-model framework for benchmarking the performance of quantum {Hamiltonian} simulations,'' \emph{IEEE Transactions on Quantum Engineering}, vol.~6, pp. 1--26, 2025.

\bibitem{Li2024DDSim}
S.~Li, Y.~Kimura, H.~Sato, and M.~Fujita, ``Parallelizing quantum simulation with decision diagrams,'' \emph{IEEE Transactions on Quantum Engineering}, vol.~5, pp. 1--12, 2024.

\bibitem{DiAdamo2021QuNetSim}
S.~DiAdamo, J.~N\"{o}tzel, B.~Zanger, and M.~M. Be\c{s}e, ``{QuNetSim}: A software framework for quantum networks,'' \emph{IEEE Transactions on Quantum Engineering}, vol.~2, pp. 1--12, 2021.

\bibitem{Stanco2022FPGA}
A.~Stanco \emph{et~al.}, ``Versatile and concurrent {FPGA}-based architecture for practical quantum communication systems,'' \emph{IEEE Transactions on Quantum Engineering}, vol.~3, pp. 1--8, 2022.

\bibitem{Xu2025MultiFPGA}
Y.~Xu, A.~D. Rajagopala, N.~Fruitwala, and G.~Huang, ``Multi-{FPGA} synchronization and data communication for quantum control and measurement,'' arXiv:2506.09856, 2025.

\bibitem{Dalvi2024Subroutine}
A.~S. Dalvi, J.~Whitlow, M.~D'Onofrio \emph{et~al.}, ``One-time compilation of device-level instructions for quantum subroutines,'' in \emph{Proc. IEEE Int. Conf. Quantum Comput. Eng. (QCE)}, Sep. 2024, pp. 873--884.

\end{thebibliography}

\end{document}